\documentclass[aps,pra,twocolumn,amsmath,amsfonts,showpacs,amssymb,floatfix,nofootinbib,groupedaddress]{revtex4}
\newcommand{\be}{\begin{eqnarray}}
\newcommand{\ee}{\end{eqnarray}}
\usepackage{graphicx,psfrag,bm,times}
\begin{document}
\title{A General Setting for Geometric Phase of Mixed States Under an Arbitrary Nonunitary Evolution}
\author{A. T. Rezakhani}
\email{rezakhani@isi.it}
\author{P. Zanardi}
\email{zanardi@isi.it} \affiliation{Institute for Scientific
Interchange (ISI), Villa Gualino, Viale Settimio Severo 65, I-10133
Torino, Italy}

\begin{abstract}
The problem of geometric phase for an open quantum system is
reinvestigated in a unifying approach. Two of existing methods to
define geometric phase, one by Uhlmann's approach and the other
 by kinematic approach, which have been considered to be distinct, are shown to be
related in this framework. The method is based upon purification of
a density matrix by its uniform decomposition and a generalization
of the parallel transport condition obtained from this
decomposition. It is shown that the generalized parallel transport
condition can be satisfied when Uhlmann's condition holds. However,
it does not mean that all solutions of the generalized parallel
transport condition are compatible with those of Uhlmann's one. It
is also shown how to recover the earlier known definitions of
geometric phase as well as how to generalize them when degeneracy
exists and varies in time.
\end{abstract}
\date{\today}
\pacs{03.65.Vf, 42.50.Dv}
\maketitle

The concept of geometric phase was originally introduced by
Pancharatnam in the classical context of comparing two polarized
light beams through their interference \cite{panch}. Later, Berry
pointed out to its importance even in quantum systems undergoing a
cyclic adiabatic evolution \cite{berry}. After that, this important
notion has been subject of interest in many different aspects, which
has led to many different generalizations and applications
\cite{anandan,shapere}. Of course in general cases to retain purely
geometrical nature of the phase one has to put some constrains,
namely parallel transport (PT) conditions. In this manner, geometric
phase is a feature which only depends on geometry of the path
traversed by the system in its motion during evolution.

 It is also worth noting
that an important source of the renewed interest in geometric phases
is their relevance to geometric quantum computation and holonomic
quantum computation \cite{zanardi}. Indeed, it is known that quantum
logic gates can be implemented only by using the concept of
geometric phases. It is believed that the purely geometric nature of
this phase makes such computations intrinsically fault tolerant and
robust against noise \cite{rob}.

A pure state is merely an idealization and in real experiments a
description of the system in terms of mixed states is usually
required. This point accounts for attempts towards extending the
concept of geometric phase to mixed states. In fact, Uhlmann was the
first in tackling the problem through the mathematical approach of
purification of mixed states \cite{uhl1}. This method is rather
general in that it is independent of the type of evolution of the
system. Next, Sj\"{o}qvist {\em et al.} put forward a quantum
interferometric based definition for geometric phase of
nondegenerate density matrices undergoing a {\em unitary} evolution
\cite{sjoq1}. Later, Singh {\em et al.} proposed a kinematic
description and extended the results to the case of degenerate mixed
states \cite{singh}. It must be mentioned that there also exists
another, differential geometric, approach to define geometric phase
for mixed states undergoing a unitary evolution \cite{chaturvedi}.
In this approach, mixed state geometric phase appears as an
immediate and direct generalization of the pure state case.

Indeed, in the case of environmental effects such as decoherence,
one has to consider nonunitary evolutions of mixed states. Some
generalizations in this direction have been addressed in Refs.~
\cite{uhl1,ericsson2,gam,pix,pati,carollo1,carollo2,nonunitary,lidar}.
The proposition in Ref.~\cite{ericsson2} for completely positive
maps in spite of being operationally well-defined depends on the
specific Kraus representation for the map. In
Refs.~\cite{carollo1,carollo2}, the problem of geometric phase of an
initially pure open quantum system, based on the standard definition
of pure state geometric phase, has been addressed through the
quantum jump method. A more recent effort is based on a kinematic
approach, with no {\em a priori} assumption about dynamics of the
system \cite{nonunitary}. However, most of these different
definitions do not agree with each other. In fact, Uhlmann's method
even in the case of unitary evolution does not agree with the
interferometric definition \cite{ericsson,slater}. The source of
such disagreement is known to be the use of different types of PT
conditions. Hence, it has been argued \cite{ericsson,slater,levay}
that these approaches are not generally equivalent and one cannot
obtain one from the other. Therefore, it could be desirable to find
a more unified approach which can bring together the previous
general ideas. Recently, in the unitary evolution case it has been
argued that using (nonorthogonal) decompositions different from
spectral decomposition can make it possible to unify the kinematic
and Uhlmann's approaches \cite{gen}. In this framework, a suitable
notion of PT condition of the mixed state is based on the PT
condition of the vectors constituting this decomposition.

In this paper, we shall use a rather similar mechanism plus uniform
decomposition of density matrices, and propose a generalized
kinematic approach for geometric phase of mixed states under an
arbitrary nonunitary evolution. This approach vividly shows how it
is possible to merge Uhlmann's approach and kinematic approach. It
is also shown how to recover the earlier definitions of geometric
phase from this more general approach. In addition, it is shown that
the approach can be easily modified to include the more general case
of degenerate mixed states. This investigation may as well be useful
in the study of robustness of geometric phases against decoherence
\cite{tidstrom}.

Let us suppose that the density matrix of our system of interest
(with the Hilbert space ${\mathcal{H}}_s$) is
$\varrho(t)=\sum_{k=1}^N p_k(t)|w_k(t)\rangle\langle w_k(t)|$, in
which $p_k(t)$'s ($|w_k(t)\rangle$'s) are considered to be its
eigenvalues (normalized eigenvectors). In a {\em general} evolution
both $p_k$ and $|w_k\rangle$ are subject to change in time. For
simplicity of our discussion, we in the sequel assume that the rank
of this matrix is constant at all instants, and even more the matrix
is nondegenerate. In the case of {\em unitary} evolution, we have
$p_k(t)=p_k(0)$ and $|w_k(t)\rangle=U(t)|w_k(0)\rangle$, where
$U(t)$ is the unitary evolution operator. However, when evolution is
nonunitary, the eigenvalues, $p_k$, can also vary in time. Thus,
generally $U(t)=\sum_k|w_k(t)\rangle\langle w_k(0)|$ does not
encompass the whole dynamical information. In fact, in such cases,
to obtain $\varrho(t)$ one often has to resort to some approximative
methods in the theory of open quantum systems \cite{open}, such as
the Lindblad equation \cite{lindblad}.

Since in our construction we use Uhlmann's PT condition \cite{uhl1}
we need to recall it briefly. Uhlmann's approach is based upon the
standard purification $\textit{w}(t)$, where
$\varrho(t)=\textit{w}(t){\textit{w}}^{\dagger}(t)$, for density
matrices. In other words, $\textit{w}$ can be considered as a
purification of $\varrho$ in the larger Hilbert space of
Hilbert-Schmidt operators with scalar product $\langle
\textit{w}(t),\textit{w}(t')\rangle=\text{tr}({\textit{w}}^{\dagger}(t)\textit{w}(t'))$
such that $\textit{w}{\textit{w}}^{\dagger}=\varrho$. It is clear
that $\textit{w}(t)=\sqrt{\varrho(t)}V(t)$ is an acceptable
purification of $\varrho$ for {\em any} unitary $V(t)$. For a
special purification where each $|\langle
\textit{w}(t),\textit{w}(t')\rangle|$ is constrained to its maximum
value Uhlmann has defined the geometric phase associated to the
evolution from $\varrho(0)$ to $\varrho(\tau)$ as
$\gamma_g(\tau)=\text{arg}(\langle
\textit{w}(0),\textit{w}(\tau)\rangle),$ where the PT condition
${\textit{w}}^{\dagger}(t)\dot{\textit{w}}(t)=\dot{\textit{w}}^{\dagger}(t)\textit{w}(t)$
 has
to be satisfied.
%\be\label{ulPT}
%${\textit{w}}^{\dagger}(t)\dot{\textit{w}}(t)=\dot{\textit{w}}^{\dagger}(t)\textit{w}(t)$.
%\ee

Let us also briefly review the construction of the geometric phase
in Ref. \cite{nonunitary}. Consider a purification for the density
matrix $\varrho(t)$ as
\be\label{purify1}&|\Psi(t)\rangle_{sa}=\sum_k\sqrt{p_k(t)}|w_k(t)\rangle_s\otimes|a_k\rangle_a,~~t\in[0,\tau].
\hskip 2mm\ee Now after imposing the PT condition;
%\be\label{PTsd}
$\langle w_k(t)|\frac{\text{d}}{\text{d}t}|w_k(t)\rangle=0$,
%\ee
 the
geometric phase, defined a la Pancharatnam \cite{panch};
$\gamma(\tau)=\text{arg}(\langle \Psi(0)|\Psi(\tau)\rangle)$, reads
as %\be\label{tongformula}
%\aligned
$\gamma_g(\tau)=\text{arg}(\sum_{k=1}^N\sqrt{p_k(0)p_k(\tau)}\langle
w_k(0)|w_k(\tau)\rangle %\\&
e^{-\int_0^{\tau}\langle w_k(t)|\dot{w}_k(t)\rangle\text{d}t})$.
Indeed, by using the PT condition one fixes the general form of the
unitary operators which like $U(t)$ can run system's dynamics. As is
clear, in this method purification of mixed state of the system is
done based on its spectral decomposition and the PT condition is
considered to be the PT condition of all the vectors constituting
this (spectral) decomposition. We know that a purification as in
Eq.~(\ref{purify1}), is only one of the possible purifications that
can give rise to the correct mixed state of the system. So, one has
the freedom to choose other decompositions and study the problem of
geometric
 phase with respect to them. In the sequel, we follow such a strategy
 and
 look for a specific purification in which all
normalized terms can be treated in a naturally uniform manner,
unlike Eq.~(\ref{purify1}) where the contribution of the $k$-th
normalized term is the time dependent variable $\sqrt{p_k(t)}$. In
other words, instead of starting from the spectral decomposition of
a density matrix which is the usual starting point of purification
based approaches, we start with another decomposition which can
result to the mentioned uniformity. In order to do so, we need the
next two important theorems
on different decompositions of a density matrix $\varrho$.\\
\indent{Theorem 1 \cite{hugh}: Let $\varrho$ has the spectral
ensemble $\{p_k,|w_k\rangle\}$. Then $\{q_l,|x_l\rangle\}$ is
another ensemble for it iff there exists a {\em unitary} matrix
${\mathcal{U}}=({\mathcal{U}}_{kl})$ such that
\be\label{th1}&\sqrt{q_l}|x_l\rangle=\sum_k\sqrt{p_k}~{\mathcal{U}}_{lk}|w_k\rangle.\ee
\indent {Theorem 2} \cite{prob}: Let $\{q_l\}$ be a probability
distribution. Then there exist normalized quantum states
$\{|x_l\rangle\}$ such that $\varrho=\sum_lq_l|x_l\rangle\langle
x_l|$, iff $\vec{q}$ is majorized by $\vec{p}$.\\
 An immediate corollary of
Theorem 2 is the existence of a {\em uniform} ensemble for {\em any}
density matrix. Therefore, there exist normalized pure states
$|x_1\rangle,\ldots,|x_{\mathcal{N}}\rangle$ such that $\varrho$ is
an equal mixture of these states with probability $1/{\mathcal{N}}$
(${\mathcal{N}}\geq N$), i.e.
$\varrho=\frac{1}{{\mathcal{N}}}\sum_{l=1}^{{\mathcal{N}}}|x_l\rangle\langle
x_l|$. For the rest of discussion we assume that ${\mathcal{N}}=N$.
Now, let us see how this uniform decomposition is related to the
spectral decomposition. By using Theorem 1, we have
%\be\label{for}
$\frac{1}{\sqrt{N}}|x_k\rangle=\sum_{l=1}^N\sqrt{p_l}~{\mathcal{U}}_{kl}|w_l\rangle$.
%\ee
It is easy to see that if one chooses an $N\times N$ Fourier matrix
(corresponding to discrete Fourier transformations \cite{fourier})
${\mathcal{U}}_{kl}=\frac{1}{\sqrt{N}}e^{-2\pi i\frac{kl}{N}}$
($k,l=0,\ldots,N-1$), and momentarily run all indices from 0 to
$N-1$ (rather than 1 to $N$) this equation is satisfied. Then, by
using a Fourier matrix one can find a uniform ensemble for any
density matrix. If we define
$C(t)=\sum_k\sqrt{p_k(t)}|w_k(0)\rangle\langle w_k(0)|$ and use the
definition of $U(t)$, we can rewrite $|x_k(t)\rangle$ in the
following matrix form
\be\label{xw}&|x_k(t)\rangle=\sqrt{N}U(t)C(t){\mathcal{U}}|w_k(0)\rangle.\ee
%where $T$ indicates transposition. Hereafter, for the ease of our
%notation we simply omit this $T$ sign.

Now we show that the above mentioned uniform decomposition is useful
in our discussion of geometric phase. Consider the following pure
state of the combined system $sa$
\be\label{purification1}%\aligned
|\Phi(t)\rangle_{sa}=\frac{1}{\sqrt{N}}\sum_k|x_k(t)\rangle_s\otimes
V(t)|a_k\rangle_a,%\\&=U(t)C(t)~{\mathcal{U}}\otimes
%V(t)\sum_k|w_k(0)\rangle_s|a_k\rangle_a,%\endaligned
\ee where $V(t)$ is the unitary evolution of the $|a_k\rangle$'s.
This state is a legitimate purification of the density matrix
$\varrho(t)$ of the system;
$\varrho(t)=\text{tr}_a(|\Phi(t)\rangle_{sa}\langle\Phi(t)|)$. If
$V(t)=I$, since $\langle x_k|x_k\rangle=1$ and all $|x_k(t)\rangle$
vectors enter with equal and constant probability of $\frac{1}{N}$
in the decomposition of the density matrix, it seems natural to
consider our (generalized) PT conditions in the form of
\be\label{xpt}\langle
x_k(t)|\frac{\text{d}}{\text{d}t}|x_k(t)\rangle=0,~~k=1,\ldots,N,\ee
that is, a density matrix undergoes a PT condition when all of the
vectors in its uniform decomposition do so. Here a point is in
order. It must be mentioned that, except the pure state case, this
PT condition is generally different from the one considered in
earlier literature \cite{sjoq1,nonunitary} \footnote{ Tong {\em et
al.}'s PT condition \cite{nonunitary} imposes a constraint on the
form of parallel transported evolution operator, $U^{\parallel}(t)$,
as $\langle
w_k(0)|U^{\parallel\dagger}\dot{U}^{\parallel}|w_k(0)\rangle=0,
~~k=1,\ldots,N,$ whereas Eq.~(\ref{xpt}) results into $\langle
w_k(0)|{\mathcal{U}}^{\dagger}CU^{\parallel\dagger}\dot{U}^{\parallel}
C{\mathcal{U}}|w_k(0)\rangle=0,~~k=1,\ldots,N.$}.

In general, in the purification (\ref{purification1}) ancillary
vectors could also vary in time, and we  have to find a natural
picture for geometric phase in this case. Let us first remind a
simple and useful property of Schmidt decomposition of bipartite
pure states \cite{gen}. If
$|\Phi\rangle_{ab}=\sum_kc_k|a_k\rangle_a|b_k\rangle_b$, then
%\be\label{schmidt}&
$(U\otimes V)|\Phi\rangle_{ab}=(UC{\mathcal{V}}^T\otimes
I)\sum_k|a_k\rangle_a|b_k\rangle_b$,
%\ee
where $C$ is a diagonal matrix in the $\{|a_k\rangle\}$ basis
defined as $C=\sum_kc_k|a_k\rangle\langle a_k|$ and
${\mathcal{V}}=\sum_{kk'}\langle
b_{k}|V|b_{k'}\rangle|a_k\rangle\langle a_{k'}|$. Here, for
notational purposes, we omit $T$ sign of ${\mathcal{V}}^T$. Now,
noting this property and assuming that the basis vectors of the
ancillary Hilbert space are $\{|w_k(0)\rangle\}$, one can rewrite
Eq.~(\ref{purification1}) as
\be\label{purification2}&|\Phi(t)\rangle_{sa}=\sum_k|\tilde{x}_k(t)\rangle_s\otimes|w_k(0)\rangle_a,\ee
where
$|\tilde{x}_k(t)\rangle=U(t)C(t){\mathcal{U}}{\mathcal{V}}(t)|w_k(0)\rangle$.
%This simply means that by moving ancilla's evolution to the system,
%the ancillary vector can always be kept fixed, though this does not
%mean that the ancilla's density matrix is fixed.
This purification
now results into the nonorthogonal decomposition
$\varrho(t)=\sum_k|\tilde{x}_k(t)\rangle\langle\tilde{x}_k(t)|$ for
the density matrix. Unlike the $\{|x_k(t)\rangle\}$ decomposition,
now for a general ${\mathcal{V}}$,
$\langle\tilde{x}_k(t)|\tilde{x}_k(t)\rangle$
 is not time independent and, as well, is not equal for all $k$'s. However, if we consider the normalized vectors
$|\hat{\tilde{x}}(t)\rangle=\frac{|\tilde{x}_k(t)\rangle}{||\tilde{x}_k(t)||}$
it still looks natural to consider our generalized PT condition to
be in the following form
 \be\label{lastpt}\langle\hat{\tilde{x}}_k(t)|\frac{\text{d}}{\text{d}t}|\hat{\tilde{x}}_k(t)\rangle=0.\ee
In terms of $|\tilde{x}_k(t)\rangle$ vectors this is equal to %\be\label{explicitPT1}
$\langle\tilde{x}_k(t)|\frac{\text{d}}{\text{d}t}|\tilde{x}_k(t)\rangle=
\frac{1}{2}\frac{\text{d}}{\text{d}t}(\langle\tilde{x}_k(t)|\tilde{x}_k(t)\rangle)$,
%\ee
 or equivalently in more detail it is \be\label{explicitPT}\aligned \langle &
w_k(0)|{\mathcal{V}}^{\dagger}{\mathcal{U}}^{\dagger}CU^{\dagger}\dot{U}C{\mathcal{U}}{\mathcal{V}}+{\mathcal{V}}^{\dagger}
{\mathcal{U}}^{\dagger}C\dot{C}{\mathcal{U}}{\mathcal{V}}+{\mathcal{V}}^{\dagger}{\mathcal{U}}^{\dagger}C^2{\mathcal{U}}\\&\times
\dot{{\mathcal{V}}}|w_k(0)\rangle=\frac{1}{2}\frac{\text{d}}{\text{d}t}(\langle
w_k(0)|{\mathcal{V}}^{\dagger}{\mathcal{U}}^{\dagger}C^2{\mathcal{U}}{\mathcal{V}}|w_k(0)\rangle).\endaligned
\ee Now let us see what is the form of Uhlmann's PT condition. We
note that $\textit{w}(t)$ operator reads as
$\textit{w}(t)=U(t)C(t){\mathcal{U}}{\mathcal{V}}(t)$. Hence, the
explicit form of Uhlmann's PT condition is
\be\aligned\label{explicitul}&{\mathcal{V}}^{\dagger}{\mathcal{U}}^{\dagger}CU^{\dagger}\dot{U}C{\mathcal{U}}{\mathcal{V}}
+{\mathcal{V}}^{\dagger}
{\mathcal{U}}^{\dagger}C\dot{C}{\mathcal{U}}{\mathcal{V}}+{\mathcal{V}}^{\dagger}{\mathcal{U}}^{\dagger}C^2{\mathcal{U}}
\dot{{\mathcal{V}}}\\
&\quad={\mathcal{V}}^{\dagger}{\mathcal{U}}^{\dagger}C\dot{U}^{\dagger}UC{\mathcal{U}}{\mathcal{V}}+{\mathcal{V}}^{\dagger}
{\mathcal{U}}^{\dagger}\dot{C}C{\mathcal{U}}{\mathcal{V}}+\dot{{\mathcal{V}}}^{\dagger}{\mathcal{U}}^{\dagger}C^2{\mathcal{U}}
{\mathcal{V}}.
\endaligned\ee
As is seen \textsc{lhs} of this equation is exactly the expression
within bra-ket of the PT condition (\ref{explicitPT}). If sandwiched
between $\langle w_k(0)|$ and $|w_k(0)\rangle$,
Eq.~(\ref{explicitul}) gives rise to
\be\label{dN}\aligned\text{\textsc{lhs} of
(\ref{explicitPT})}&=\frac{1}{2}\langle
w_k(0)|\textsc{lhs}+\text{\textsc{rhs} of
(\ref{explicitul})}|w_k(0)\rangle\\&=\frac{1}{2}\frac{\text{d}}{\text{d}t}(\langle
w_k(0)|{\mathcal{V}}^{\dagger}{\mathcal{U}}^{\dagger}C^2{\mathcal{U}}{\mathcal{V}}|w_k(0)\rangle)\hskip
2mm.\endaligned\ee This is what we wanted to show; by using
Uhlmann's PT condition the generalized PT conditions
(\ref{explicitPT}) are also satisfied. %Up to now the two PT
%conditions have been linked to each other.
 However, it must be noted
that generally number of equations of the two PT conditions are not
equal. In other words, Eq.~(\ref{explicitul}) is a matrix equation
which constitutes $N^2$ different equations (for ${\mathcal{V}}$)
though Eq.~(\ref{lastpt}) is just a set of $N$ equations. This
simply means that there might be solutions of Eq.~(\ref{explicitPT})
that are not solutions of Eq.~(\ref{explicitul}). If it is assumed
that ${\mathcal{V}}(t)=e^{-i\tilde{H}(t)}$, then solution of
Eq.~(\ref{explicitul}) is as follows
\cite{uhl1}\be\label{ulsol}\aligned
-i\tilde{H}(t)=&-2\sum_{kk'}{\mathcal{U}}^{\dagger}|w_{k'}(0)\rangle\langle
w_k(0)|{\mathcal{U}}\\&\times\int_0^t {\text{d}}t'\langle
w_{k'}(t')|\dot{w}_k(t')\rangle
\frac{\sqrt{p_{k'}(t')p_k(t')}}{p_{k'}(t')+p_{k}(t')}.\endaligned\ee
Now it is easy to show that Eq.~(\ref{explicitPT}) can have
solutions other than (\ref{ulsol}). For example, if we suppose that
$[{\mathcal{U}}{\mathcal{V}},C]=0$, %that is
%${\mathcal{U}}{\mathcal{V}}$ and $C$ have common eigenvectors%
 and
${\mathcal{U}}{\mathcal{V}}=\sum_k e^{-il_k(t)}|w_k(0)\rangle\langle
w_k(0)|$, then Eq.~(\ref{explicitPT}) gives \be\label{li1} &
l_k(t)=-i\int_0^t\text{d}t'\langle w_k(t')|\dot{w}_k(t')\rangle,\ee
which does not generate a ${\mathcal{V}}(t)$ compatible with
(\ref{ulsol}). This comes from the fact that to satisfy
Eq.~(\ref{explicitPT}) we only need to have the diagonal terms of
Uhlmann's PT condition, whereas off-diagonal terms of this equation
may put extra constraints that are redundant for validity of
Eq.~(\ref{explicitPT}).%% So we are left with a freedom to choose
%other off-diagonal terms with Eq.~(\ref{explicitPT}). %Then solutions
%of the both equations only must have equal diagonal elements.

Now geometric phase can be simply defined a la Pancharatnam as
\be\label{geom} \gamma_g(t)&=\text{arg}(\langle
\Phi(0)|\Phi(t)\rangle)=\text{arg}(\sum_k
\nu_k(t)e^{i\gamma_k(t)}),\ee where $\langle
\tilde{x}_k(0)|\tilde{x}_k(t)\rangle=\nu_k(t)e^{i\gamma_k(t)}$, i.e.
$\nu_k(t)$ ($\gamma_k(t)$) is the visibility (geometric phase) of
the $k$-th component of $|\Phi(t)\rangle$. The explicit form is
obtained by insertion of the definition of $|\tilde{x}_k(t)\rangle$
in this equation, which gives
\be\label{}\aligned\gamma_g(t)=&\text{arg}(\sum_{kk'}\sqrt{p_{k'}(0)p_k(t)}\langle
w_{k'}(0)|w_k(t)\rangle\\&\times\langle
w_k(0)|{\mathcal{U}}{\mathcal{V}}(t){\mathcal{U}}^{\dagger}|w_{k'}(0)\rangle).\endaligned\ee
This equation shows that geometric phase, as described here to be
combined with Uhlmann's definition, generally retains a memory of
the evolution of both system and the ancilla, that is, it is a
general property of the whole system which depends on the history of
the system as well as the history of the ancilla entangled with it
\cite{ericsson}.

In the remainder, we investigate how the earlier definitions of
geometric phase \cite{sjoq1,nonunitary} can be obtained from the
present framework as special cases. If we confine ourselves to a
restriction of the solution of (\ref{ulsol}) for
${\tilde{\mathcal{V}}}(t)$, such that
${\tilde{\mathcal{V}}}_{kk'}(t)={\mathcal{V}}_{kk'}(t)\delta_{kk'}$
and has the property
\be\label{commutation}&[\tilde{{\mathcal{V}}},{\mathcal{U}}^{\dagger}C^2{\mathcal{U}}]=0,\ee
or equivalently
%\be\label{Vform}&
$\tilde{{\mathcal{V}}}(t)=\sum_ke^{-il_k(t)}{\mathcal{U}}^{\dagger}|w_k(0)\rangle\langle
w_k(0)|{\mathcal{U}}$,
%\hskip 2mm\ee
where $l(t)$ is defined as in Eq.~(\ref{li1}), then explicit form of
$\gamma_g$ becomes \be\hskip
-3mm&\gamma_g(t)=\text{arg}(\sum_k\sqrt{p_k(0)p_k(t)}\langle
w_k(0)|w_k(t)\rangle e^{-il_k(t)}),\hskip 3mm\ee as in Ref.
\cite{nonunitary}. Thus, in the context of the discussion of
Ref.~\cite{ericsson2}, it can be said that the physical role of the
commutation relation (\ref{commutation}) appears like removing
memory effects of ancilla's evolution from geometric phase.

Let us end by mentioning some remarks on the initial assumptions of
the approach while our stress is still on derivation of earlier
results and their possible generalizations.
%It is shown that the
%restrictive assumptions can be easily relaxed, and in this sense the
%approach is general.
Based upon Theorem 2, it is seen that one can always choose
${\mathcal{N}}$, number of the vectors in uniform decomposition,
such that ${\mathcal{N}}\geq N$. %Thus, when $N$, rank of the density
%matrix, is a time varying parameter,
%in order to have Eq.~(\ref{aq1})
For example, we can assume that
${\mathcal{N}}=\text{dim}({\mathcal{H}}_s)$. %, which is a constant.
Now we show how the whole framework can be modified in the
degenerate case. Consider the evolution for the density matrix of
the system %\be\label{deg1}&
from $\varrho(0)$ to
$\varrho(t)=\sum_{k=1}^N\sum_{\mu=1}^{n_k}p_k(t)|w_k^{\mu}(t)\rangle\langle
w_k^{\mu}(t)|$, %\ee
where $p_k(t)$, $k=1,\ldots,N$, are the
$n_k$-fold degenerate eigenvalues of $\varrho(t)$, and
$|w_k^{\mu}(t)\rangle$, $\mu=1,\dots,n_k$, are considered the
corresponding eigenvectors. In this case, the pure state of the
total system is
%\be\label{deg2}&
$|\Phi(t)\rangle_{sa}=\sum_{k=1}^N\sum_{\mu=1}^{n_k}|\tilde{x}_k^{\mu}(t)\rangle_s\otimes
|w_{k}^{\mu}(0)\rangle_a$,
%\ee
where $|\tilde{x}_k^{\mu}(t)\rangle$ is
defined as in Eq.~(\ref{purification2}) in which $|w_k(0)\rangle$ is
replaced by $|w_k^{\mu}(0)\rangle$. Then, one notes that
\be\label{deg3}\aligned\langle\Phi(0)|\Phi(t)\rangle=&\sum_{kk'\mu\mu'}\sqrt{p_k(0)p_{k'}(t)}\langle
w_k^{\mu}(0)|w_{k'}^{\mu'}(t)\rangle\\&\quad\times\langle
w_k^{\mu}(0)|{\mathcal{U}}{\mathcal{V}}(t){\mathcal{U}}^{\dagger}|w_{k'}^{\mu'}(0)\rangle,
\endaligned\ee
which is determined when all elements of ${\mathcal{V}}(t)$ are
known. Now we choose our PT condition in this general case as
\be\label{lastPT}\langle\hat{\tilde{x}}_k^{\mu}(t)|\frac{\text{d}}{\text{d}t}
|\hat{\tilde{x}}_k^{\mu'}(t)\rangle=0,~~~~\mu,\mu'=1,\dots,n_k.\ee
It can be checked that this PT condition can also be satisfied by
assuming Uhlmann's PT condition,
Eq.~(\ref{explicitul}). %Since
%${\mathcal{U}}C(t){\mathcal{U}}^{\dagger}$ now has degeneracy, by
%using its commutativity with $\tilde{{\mathcal{V}}}$, one cannot
%obtain form of $\tilde{{\mathcal{V}}}$ as before.
 In this case, it
is easily seen that the most general form for
$\tilde{{\mathcal{V}}}$ which satisfies Eq.~(\ref{commutation}) is
as follows
\be\label{Vformdeg}&\tilde{{\mathcal{V}}}(t)=\sum_{k\mu\mu'}\alpha_k^{\mu\mu'}(t)~
{\mathcal{U}}^{\dagger}|w_k^{\mu}(0)\rangle\langle
w_k^{\mu'}(0)|{\mathcal{U}}.\ee  After some algebra and using the
commutation relation (\ref{commutation}), it is obtained that
%\be\label{deg5} %\aligned&\langle
%w_k^{\mu}(0)|{\mathcal{U}}\tilde{{\mathcal{V}}}(t){\mathcal{U}}^{\dagger}|w_k^{\mu'}(0)\rangle\\&\quad
$\alpha_k^{\mu\mu'}(t)= \langle
w_k^{\mu}(0)|\textbf{P}e^{-\int_0^tU^{\dagger}(t')\dot{U}(t')\text{d}t'}|w_k^{\mu'}(0)\rangle$,
%\endaligned
%\ee
where \textbf{P} denotes path ordering. After inserting this
relation back
into Eq.~(\ref{deg3}), non-Abelian factors show up in the geometric phase.%, as expected in the
%degenerate case \cite{nonunitary,zanardi}.

In general, when degeneracies vary in time, a level--crossing like
behavior can occur. In this situation, in the discussion of
differentiability of the eigenvalues (and eigenvectors) the notion
of ordering of the eigenvalues becomes important. For example, it
can happen that the natural ordering as $p_1(t)\geq\dots\geq p_N(t)$
(for all $t$) destroys differentiability, thus, one has to seek for
some ordering which respects it \cite{Bhatia}. If such an ordering
can be found, then the operator $U(t)$, eigenvalues, and
eigenvectors are still well--defined differentiable functions and
our approach may be generalized as well.

In summary, the notion of geometric phase of a mixed state
undergoing nonunitary evolution has been investigated in a unifying
picture in which two of the previous general definitions, Uhlmann's
definition and kinematic approach, have been related to each other.
In this formalism, we have used the idea of purification of state of
a system by uniform decomposition of its density matrix rather than
the spectral one, and by attaching a time varying ancilla to it.
Then, as a natural choice for parallel transport condition, we have
considered that a mixed state is undergoing parallel transport
condition when all the (normalized) vectors of its corresponding
purification are subject to this condition. This generalized
parallel transport condition is different from the ones defined
previously in the literature. It has been shown that the new
conditions are satisfied when Uhlmann's condition holds. However,
because of different numbers of equations in the two parallel
transport conditions, the generalized parallel conditions are only
diagonal equations of Uhlmann's condition. Finally, it has been
shown how to recover earlier definitions of geometric phase of a
mixed state. Extension of the method to the more general cases of
degenerate density matrices with %nonconstant rank and
time varying
degeneracies have also been discussed.

We thank N. Paunkovic for useful discussions. This work was
supported by EU project TOPQIP under Contract No. IST-2001-39215.

%%%%%%%%%%%%%%%%%%%%%%%%%%%%%%%%%%%%%%%%%%%%%%%%%%%%%%%%%%%%%%

%%%%%%%%%%%%%%%%%%%%%%%%%%%%%%%%%%%%%%%%%%%%%%%%%%%%
\end{document}